\documentclass[12pt]{article}

\pdfoutput=1

\usepackage{amsmath,amssymb,amsfonts}
\usepackage{color}
\definecolor{darkblue}{rgb}{0.1,0.1,.7}
\usepackage[colorlinks, linkcolor=darkblue, citecolor=darkblue, urlcolor=darkblue, linktocpage]{hyperref} 
\usepackage[square, comma, compress,numbers]{natbib}
\usepackage[]{amsmath}
\usepackage[]{graphicx}
\usepackage[]{latexsym}
\usepackage{geometry}
\usepackage{amscd}
\usepackage[all,cmtip]{xy}
\usepackage{mathrsfs}
\usepackage[margin=10pt,font=small,labelfont=bf]{caption}
\numberwithin{equation}{section}

\newcommand{\reef}[1]{(\ref{#1})}

\newcommand{\be}{\begin{equation}}
\newcommand{\ee}{\end{equation}}
\newcommand{\bea}{\begin{eqnarray}}
\newcommand{\eea}{\end{eqnarray}}

\def\beq{\begin{equation}} 
\def\eeq{\end{equation}} 

\def\del {\partial} 
\def\nn{\nonumber} 
 
\def\bR {\mathbb{R}} 
\def\calO {{\cal O}}

\def\bn{{\mathbf{n}}}

\def\ge{\geqslant}
\def\le{\leqslant}

\def\T{\beta}

\begin{document}

\vspace*{-.6in} \thispagestyle{empty}
\begin{flushright}
LPTENS--12/31
\end{flushright}
\vspace{.2in} {\Large
\begin{center}
{\bf OPE Convergence in Conformal Field Theory}
\end{center}
}
\vspace{.2in}
\begin{center}
{\bf 
Duccio Pappadopulo$^{a}$, Slava Rychkov$^{b}$, Johnny Espin$^{a}$, Riccardo Rattazzi$^{a}$
}
\\
\vspace{.6in} 
$^a$ {\it Institut de Th\'eorie des Ph\'enom\`enes Physiques, EPFL, CH-1015 Lausanne, Switzerland}
\\\vspace{.2in} 
$^b$ {\it Facult\'{e} de Physique, Universit\'{e} Pierre et Marie Curie\\
\& Laboratoire de Physique Th\'{e}orique, \'{E}cole Normale Sup\'{e}rieure, Paris, France}
\end{center}

\vspace{.2in}

\begin{abstract}
We clarify questions related to the convergence of the OPE and conformal block decomposition in unitary Conformal Field Theories
(for any number of spacetime dimensions). In particular, we explain why these expansions are convergent in a finite region.
We also show that the convergence is exponentially fast, in the sense that the operators of dimension above $\Delta$
contribute to correlation functions at most $\exp(-a\Delta)$. Here the constant $a>0$ depends on the positions of operator 
insertions and we compute it explicitly. 

\end{abstract}
\vspace{.2in}
\vspace{.3in}
\hspace{0.7cm} August 2012

\newpage

\tableofcontents

\section{Introduction}
\label{sec:intro}

Conformal Field Theories (CFTs) appear in many branches of theoretical physics: fundamental interactions beyond the Standard Model, string theory and quantum gravity, inflationary cosmology etc. They play a particularly important role in all problems involving renormalization group (RG) flows, such as the theory of critical phenomena. Indeed, a general Quantum Field Theory (QFT) can be thought of as an RG flow starting from a CFT in the UV and flowing to another CFT in the IR.

CFTs are also under better theoretical control than general QFTs, with several tools which are not available or much less powerful without conformal symmetry. In this paper we will be concerned with one such tool: the Operator Product Expansion (OPE).

For simplicity we will be considering here the OPE of two identical scalar operators. In any Lorentz-invariant QFT, the operator product $\phi(x)\phi(0)$ in the short-distance limit $x\to0$ can be approximated by a sum of local operators with $c$-number coefficient functions depending only on $x^2$:
\beq
\phi(x)\phi(0)\stackrel{x\to 0}{\simeq} \sum _{O} C(x^2) x^{\mu_1}\ldots x^{\mu_l} O_{\mu_1\ldots\mu_l}(0)\,.
\label{eq:OPEgen}
\eeq
The RHS will in general contain both scalars and operators of nonzero spin $l$, and in the latter case their indices have to be contracted as shown.

In the conformal case, Eq.~\reef{eq:OPEgen} has three more specific properties. First of all, the $x^2$ dependence of the coefficient functions is a power law fixed in terms of the operator scaling dimensions\footnote{Strictly speaking, this is only guaranteed in a unitary theory so that the dilatation operator can be diagonalized. There is a special class of non-unitary theories called logarithmic CFTs in which the dilatation operator has Jordan blocks, and some correlation function contain logarithmic deviations from power-laws. In this paper we only consider unitary theories.}
\beq
C(x^2)=const. (x^2)^{-\Delta_\phi+(\Delta_{\mathcal{O}}-l)/2}\,.
\eeq

Second, conformal symmetry allows us to classify all local operators of the theory into primaries, which transform homogeneously under the conformal group,\footnote{$\equiv$ the group $SO(D+1,1)$ of global conformal transformations of the $D$-dimensional Euclidean space. Our main results will apply for any $D$, although for $D=2$ stronger results could be obtained using the full Virasoro algebra. What we call primaries is usually called quasi-primaries in 2D.} and their derivatives (descendants). 
We can now rewrite the OPE as a summation just over primaries, denoted by curly $\calO$:
\beq
\phi(x)\phi(y)={\sum _{\mathcal{O}}} f_{\phi\phi\mathcal{O}}P(x-y,\del_y) \calO(y)\,.
\label{eq:cOPE}
\eeq
The index contractions if $\calO$ has nonzero spin are implicit. The coefficient function $P$ is now a power series in $\del_y$ and encodes the contribution of $\calO$ and all of its descendants. Up to the overall numerical coefficient $\lambda_{\phi\phi\calO}$, the form of this function is completely fixed by conformal symmetry in terms of the operator scaling dimensions. 

To make this more concrete, let us discuss the case of a scalar $\calO$ in some detail. Consider the three point function $\langle\phi\phi\calO\rangle$. On the one hand it is fixed by conformal symmetry:
\begin{align}
\langle\phi(x)\phi(y)\calO(z)\rangle&=f_{\phi\phi\calO}|x-y|^{\Delta_\calO-2\Delta_\phi}|x-z|^{-\Delta_\calO}|y-z|^{-\Delta_\calO}\,.
\label{eq:3ptsc}
\end{align}
On the other hand according to the OPE \reef{eq:cOPE} it should be computable as:
\beq
\langle\phi(x)\phi(y)\calO(z)\rangle=f_{\phi\phi\calO}P(x-y,\del_y) \langle\calO(y)\calO(z)\rangle
\label{eq:OPE3pt}\,,
\eeq
where
\beq
\langle\calO(y)\calO(z)\rangle= |y-z|^{-2\Delta_\calO}\,.
\label{eq:2pt}
\eeq
Demanding that the two expressions agree, the coefficient function $P$ can be fixed unambiguously. Here are the first few terms in the expansion:
\begin{gather}
P(x,\del_y)=|x|^{\Delta-2\Delta_\phi}\Bigl[1+\frac 12 x^\mu \del_\mu
+\alpha x^\mu x^\nu\del_\mu\del_\nu+\beta x^2 \del^2+\ldots\Bigr]\,,
\label{eq:3t}\\
\alpha=\frac{\Delta+2}{8(\Delta+1)}\,,\qquad \beta=-\frac{\Delta}{16(\Delta-D/2+1)(\Delta+1)}\,.\nn
\end{gather}
For the OPE of two identical scalars considered here the coefficients depend on $\Delta\equiv\Delta_\calO$ but not on $\Delta_\phi$; in general they would also depend on $\Delta_{\phi_1}-\Delta_{\phi_2}$. Such expansions have been worked out to all orders (and also for $\calO$ of nonzero spin) in the 1970's \cite{Ferrara:1971zy,Ferrara:1971vh,Ferrara:1972cq} and show interesting structure visible already in \reef{eq:3t}.
For example, the $D$ dependence appears only in the terms multiplied by $x^2$, and would therefore be subleading on the light cone. Notice also that the $D$-dependent term becomes singular when $\Delta_\calO$ hits the scalar field unitarity bound $D/2-1$. This is not a problem since such an $\calO$ is necessarily free and so $f_{\phi\phi\calO}=0$.

Once the OPE structure is determined, we can use it to express any $n$-point function as a sum of $(n-1)$-functions. Schematically:
\beq
\langle\phi(x)\phi(y)\prod\psi_i(z_i)\rangle={\sum_{\calO}} f_{\phi\phi\calO}P(x-y,\del_y) \langle\calO(y)
\prod\psi_i(z_i)\rangle\,.
\label{eq:OPEnpt}
\eeq
For $n=3$ there is a single exchanged primary $\calO=\psi_1$, and we go back to Eq.~\reef{eq:OPE3pt}, but for $n\ge 4$ the sum will be infinite. Actually, it will be doubly infinite since $P$'s are infinite series in $\del_y$.

And here comes the third special property of the conformal OPE: \emph{it converges}. By this we mean that the representations \reef{eq:OPEnpt} are actually absolutely convergent at finite separation $x-y$, rather than being just asymptotic expansions in the limit $x\to y$.

This property has two important applications:
\begin{itemize}
\item 
Correlation functions of arbitrarily high order can be computed by applying the OPE recursively. Of course, to do this we must know all primary operator dimensions $\Delta_i$ and all OPE coefficients $f_{ijk}$ (collectively known as the \emph{CFT data}).
\item
Eq.~\reef{eq:OPEnpt} can also be used to constrain the CFT data itself, by means of an old idea known as the conformal bootstrap \cite{Ferrara:1973yt,Polyakov:1974gs,Belavin:1984vu}. The point is that a conformal four point function can be computed using the OPE in three different channels: (12)(34), (13)(24), and (14)(23). That the results agree is a constraint on the CFT data. 
\end{itemize}

To explain how this works in the simplest setting, consider the correlator of four identical scalars
\beq
\langle\phi(x_{1})\phi(x_{2})\phi(x_{3})\phi(x_{4})\rangle
=\frac{g(u,v)}{x^{2\Delta_\phi}_{12}x_{34}^{2\Delta_\phi}}\qquad(x_{ij}\equiv x_i-x_j)\,, 
\label{eq:4pteq}%
\eeq
constrained by the conformal symmetry to have this form with $g(u,v)$ a function of the cross ratios
\begin{equation}
u=\frac{x_{12}^{2}x_{34}^{2}}{x_{13}^{2}x_{24}^{2}}\,,\quad v=\frac{x_{14}%
^{2}x_{23}^{2}}{x_{13}^{2}x_{24}^{2}}\,.
\end{equation}
The same four point function can be computed by using the OPE twice in the (12)(34) channel, which gives:
\beq
\langle\phi\phi\phi\phi\rangle
={\sum_\calO}(f_\calO)^2P(x_1-x_2,\del_2)P(x_3-x_4,\del_4)\langle\calO(x_2)\calO(x_4)\rangle
\equiv{\sum_\calO}(f_\calO)^2\frac{G_\calO(u,v)}{x^{2\Delta_\phi}_{12}x_{34}^{2\Delta_\phi}}\,.
\label{eq:cbdef}
\eeq
The \emph{conformal blocks} $G_\calO(u,v)$ defined by the last equality (see \cite{Costa:2011dw} for a longer introduction) are completely fixed by conformal symmetry; they depend on $\calO$'s dimension and spin as parameters. The function $g(u,v)$ can thus be expressed as a series in conformal blocks:
\beq
g(u,v)=1+{\sum_{\mathcal{O}}} (f_{\mathcal{O}})^2 G_{\mathcal{O}}(u,v)\,,
\label{eq:CBdeceq}
\eeq
where the first term represents the contribution of the unit operator in the OPE $\phi\times\phi$. 

Eq.~\reef{eq:CBdeceq} solves the problem of computing the four point function in terms of the CFT data. To derive this equation, we made an arbitrary choice of applying the OPE in the (12)(34) channel. However, the representation in the crossed channel (14)(23) should be equally valid. In the considered case of four identical scalar, the crossed and direct channels involve the same primary exchanges, and their equivalence can be expressed as a functional equation satisfied by the function $g(u,v)$:
\beq
g(u,v)=(u/v)^{\Delta_\phi}g(v,u)\,.
\eeq
Substituting the conformal block expansion we get
\beq
[v^{\Delta_\phi}-u^{\Delta_\phi}]+{\sum_{\mathcal{O}}} (f_{\mathcal{O}})^2 [v^{\Delta_\phi}G_{\mathcal{O}}(u,v)-u^{\Delta_\phi}G_{\mathcal{O}}(v,u)]=0\,.
\label{eq:cross-sum}
\eeq
Since 1970's it was hoped that this equation and its generalizations to more general four point functions impose nontrivial constraints on the CFT data. In the last few years this hope is starting to be realized. A practical way of extracting information from Eq.~\reef{eq:cross-sum} is as follows \cite{Rattazzi:2008pe} (see \cite{Rychkov:2009ij,Caracciolo:2009bx,Poland:2010wg,Rattazzi:2010gj,Rattazzi:2010yc,Vichi:2011ux,Poland:2011ey,ElShowk:2012ht} for generalizations, and \cite{Heemskerk:2009pn} for a different method used for large $N$ theories). 

\begin{figure}[htbp]
\begin{center}
\includegraphics[width=6cm]{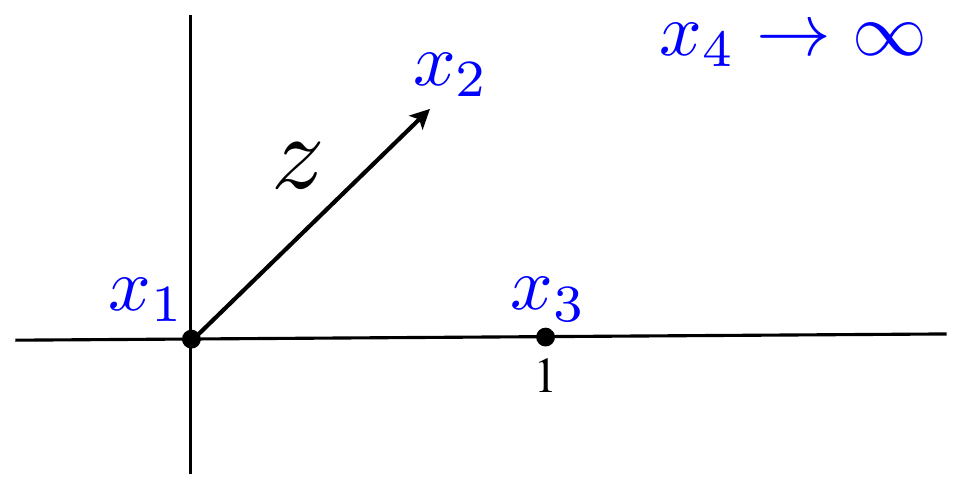}
\caption{Using conformal freedom, three operators can be fixed at $x_1=0$, $x_3=(1,0,\ldots,0)$, $x_4\to \infty$, while the fourth point  $x_2$ can be assumed to lie in the (12) plane. The variable $z$ is then the complex coordinate of $x_2$ in this plane, while $\bar z$ is its complex conjugate. }
\label{fig:z}
\end{center}
\end{figure}

Using conformal freedom, let's put the four operator insertions in a plane with three points fixed at 0, 1 and $\infty$ (see Fig.~\ref{fig:z}). Conformal blocks are then functions of the fourth point complex coordinate $z$, and the cross ratios are given by
\beq
u=z\bar{z}\,,\qquad v=(1-z)(1-\bar z)\,.
\label{eq:zzbar}
\eeq
Eq.~\reef{eq:cross-sum} is then Taylor-expanded around the point $z=1/2$, and each coefficient is equated to zero. This gives a system of linear equations for infinitely many unknown non-negative coefficients $p_\calO\equiv(f_{\mathcal{O}})^2$.

One can now try to make an assumption about the spectrum, like that there is no scalar operator below a certain dimension, or that there is a gap between the stress tensor and the next spin two operator, etc. Any such assumption can be encoded by saying that $p_{\mathcal{O}}$ should vanish in a certain interval. One then asks if there are nontrivial solutions satisfying the assumption, which is a question of linear algebra and can be decided numerically. If there are none, then a CFT with such a spectrum cannot exist. 

For reasons which are not yet fully understood, this simple-minded approach gives pretty strong, sometimes optimal \cite{Rychkov:2009ij,ElShowk:2012ht}, bounds on the operator spectrum and also on the OPE coefficients. Partly this success must be due to the fortunate choice of the Taylor-expansion point $z=1/2$, halfway between the direct channel OPE limit $z\to0$ and the crossed channel limit $z\to1$. But does the conformal block expansion converge near this point, as is necessary for the validity of the method? We noticed on several occasions that it does, supporting this statement with brief arguments of various convincing power. However, the issue is sufficiently important to warrant a detailed investigation. This is the main purpose of this paper.

The discussion will be structured as follows. We begin in section \ref{sec:pol} by sketching a well-known argument relating convergence of the OPE expansion to the convergence of a scalar product of two Hilbert space states in the radial quantization. We believe that this way of viewing the problem is correct, and will try to eliminate whatever doubts may still exist in its validity. First of all, to illustrate the Hilbert space techniques used in the argument we perform a few simple CFT computations and check that results agree with other methods (section \ref{sec:op}). The heart of the paper is section \ref{sec:4pt}, where we focus on a CFT four point function and discuss the rate (as opposed to just the region) of OPE convergence for this correlator. We show that this convergence is exponentially fast and derive the exponent. 
As a prerequisite to this result we derive asymptotics of the CFT spectral density weighted by the OPE coefficients, which is of independent interest. In section \ref{sec:cb}, we show that the convergence of the conformal block decomposition can be proven by the same methods, and that it is also exponentially fast. Consequences and potential applications of our results are discussed in section \ref{sec:disc}. 

\section{Radial quantization argument}
\label{sec:pol}
The following result concerning the convergence of the OPE is rather well known: expansion \reef{eq:OPEnpt} will converge as long as $x$ is closer to $y$ than any other operator insertion:
\beq
|x-y|<\min_i |z_i-y|\,.
\label{eq:rad}
\eeq

\begin{figure}[htbp]
\begin{center}
\includegraphics[width=5cm]{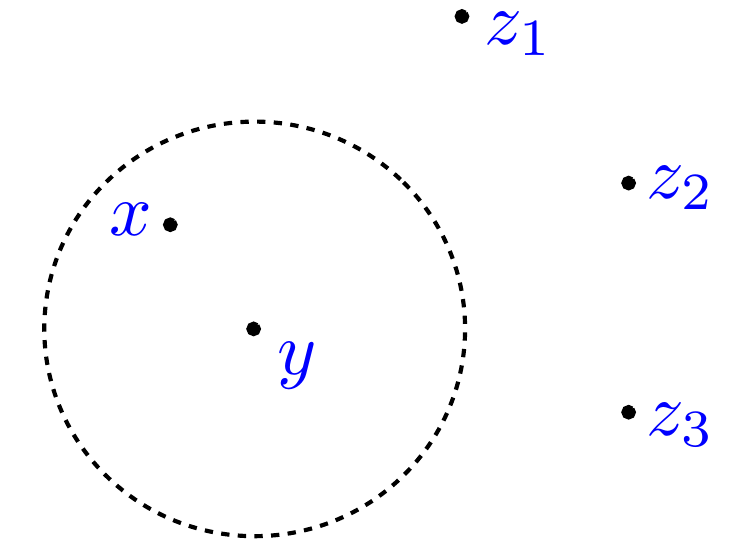}
\caption{An $n$-point correlation function can be viewed as the overlap of two states in the radial quantization Hilbert space, living on a sphere.}
\label{fig:PhiPsi}
\end{center}
\end{figure}

{\it Sketch of proof.} (see Polchinski \cite{Polchinski:1998rq}, Sec~2.9) Let's quantize the theory radially with the point $y$ as the origin. If the condition \reef{eq:rad} is satisfied, we can find a sphere separating the points $x,y$ from the rest of the operators (see Fig.~\ref{fig:PhiPsi}). The correlation function in the LHS of \reef{eq:OPEnpt} can then be viewed as the overlap
\beq
\langle \Psi|\Phi\rangle
\eeq
between the two states living on this sphere which are produced by acting with $\phi$'s and $\psi$'s on the radial quantization \emph{in} and \emph{out} vacua:
\beq
|\Phi \rangle=\phi(x)\phi(y)|0\rangle\,,\qquad \langle \Psi|=\langle 0|\prod\psi(z_i)\,.
\eeq
Furthermore, the state $|\Phi \rangle$ can be expanded into a complete basis of energy eigenstates: 
\beq
|\Phi \rangle = \sum_n C_n(x-y) |E_n\rangle\,.
\label{eq:stateexp}
\eeq
Since the radial quantization Hamiltonian is the dilatation generator, these states are generated by acting on the vacuum by local operators of definite scaling dimension $\Delta_n=E_n$. Moreover, there is one-to-one correspondence between this expansion and 
the OPE \reef{eq:cOPE}. For every primary $\calO$ in the OPE, the expansion \reef{eq:stateexp} will contain a series of states produced by $\calO(y)$ and all of its descendants. Schematically:
\beq
|E_n\rangle =(\del_y)^n \calO(y)|0\rangle\,\qquad E_n=\Delta_\calO+n\,.
\label{eq:states}
\eeq
The coefficients $C_n(x-y)$ with which such a state enters into the expansion is found by simply picking up the coefficient of $(\del_y)^n$ in the conformal OPE. 

Convergence of the OPE \reef{eq:OPEnpt} now follows from a basic theorem about Hilbert spaces: {\it the scalar product of two states converges when one of the two states is expanded into an orthonormal basis.}  Q.E.D.

The above argument will provide a starting point for our discussion. Our goal will be to make it more explicit and quantitative, in particular by determining the rate of convergence.

\section{Operator formalism exemplified}
\label{sec:op}

The purpose of this section is to provide some background material about the radial quantization and the state-operator correspondence. This is pretty standard and may be skipped by the experts.

\subsection{Map to the cylinder}
\begin{figure}[htbp]
\begin{center}
\includegraphics[height=5cm]{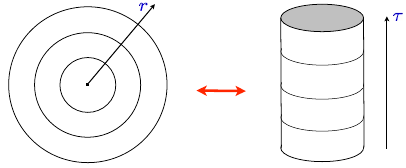}
\caption{The map between $\bR^D$ and the cylinder.}
\label{fig:map}
\end{center}
\end{figure}

One way to think about the radial quantization is by mapping the CFT from the Euclidean flat $D$-dimensional space to the cylinder $\bR\times S^{D-1}$ (Fig.~\ref{fig:map}). In the 2D case this is usually carried out by means of the logarithmic coordinate transformation. However, the map exists in any $D$ because the cylinder is conformally flat:
\beq
ds^2_\text{cyl}=d\tau^2+d\mathbf{n}^2=r^{-2}(dr^2+r^2 d\mathbf{n}^2)\equiv r^{-2} ds^2_{\bR^D}\qquad (\tau=\log r\,,\bn^2=1)\,.
\eeq
On the cylinder we have time translation invariance and the usual Hamiltonian quantization. Going back to the flat space the cylinder time slicing is mapped onto the slicing by spheres, and we recover the radial quantization. 

It is a basic property of CFT that correlation functions on conformally flat backgrounds can be computed by rescaling the flat space correlation functions. For scalars, we have:\footnote{See e.g.~\cite{Cardy:1988cwa,Klebanov:2011td} for specific applications of this rule. The only exception from the rule is the stress tensor Weyl anomaly \cite{Cappelli:1988vw}.} 
\beq
\langle \phi(x)\ldots\rangle_{e^{2\sigma(x)} dx^2}=e^{-\sigma(x)\Delta_\phi}\langle \phi(x)\ldots\rangle_{\bR^D}\,,
\eeq
which on the cylinder becomes:
\beq
\phi(\tau,\bn)_\text{cyl}=r^{\Delta_{\phi}}\phi(x)_{\bR^D}\,.
\label{eq:cylrd}
\eeq

Let us demonstrate this rule by computing the two point function on the cylinder starting from the flat space expression \reef{eq:2pt}. We get
\cite{Luscher:1974ez}:
\beq
\langle \calO(\tau_2,\mathbf{n}_2) \calO(\tau_1,\mathbf{n}_1)\rangle_\text{cyl} 
=e^{-\T\Delta } |1-2e^{-\T}(\bn_1\cdot\bn_2)+e^{-2\T}|^{-\Delta}\,.
\label{eq:luscher}
\eeq
The cylinder times enter only via their difference $\T=\tau_2-\tau_1$, consistently with time translation invariance.
It is also interesting to expand the two point function in the large $\T$ limit:
\beq
e^{-\T\Delta} \sum_{n=0}^\infty c_n e^{-\T n}\,.
\label{eq:Geg}
\eeq
The powers of $e^{-\T}$ appearing in this expression are consistent with the tower of states \reef{eq:states} propagating along the cylinder. 
The coefficients $c_n$ can be found explicitly: they are the Gegenbauer polynomials entering via their generating function:
\beq
c_n=C_n^{(\Delta)}(\bn_1\cdot\bn_2)\,.
\label{eq:gegexpl}
\eeq 
We will reproduce this result via the operator methods below.

\subsection{Positivity and conjugation}
\begin{figure}[htbp]
\begin{center}
\includegraphics[height=4.5cm]{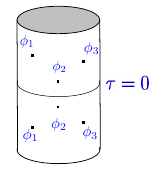}
\caption{Example of a reflection-positive six point function on the cylinder.}
\label{fig:6pt}
\end{center}
\end{figure}

We next discuss reflection positivity and conjugations. On the cylinder we have the time reflection transformation $\theta:\tau\to-\tau$. Any $(2n)$-point correlation function with operator inserted symmetrically under this transformation will be positive (in a unitary theory). In fact this correlation function can be viewed as computing the norm of a state generated by half of the operators (see Fig.~\ref{fig:6pt}) acting on the vacuum. This can also be seen in the Hamiltonian  formulation where  $\phi(\tau,\bn)_\text{cyl}\equiv e^{\tau H_\text{cyl}}\phi(0,\bn) e^{-\tau H_\text{cyl}}$, so that for an hermitean Minkowskian field $\phi(0,\bn)$ one has
\begin{equation}
\phi(\tau,\bn)_\text{cyl}^\dagger = (e^{\tau H_\text{cyl}}\phi(0,\bn) e^{-\tau H_\text{cyl}})^\dagger = e^{-\tau H_\text{cyl}}\phi(0,\bn) e^{\tau H_\text{cyl}}=\phi(-\tau,\bn)_\text{cyl}\, .
\label{eq:hamiltonian_conjugation}
\end{equation}
Reflection positivity in the Euclidean theory
 then corresponds to unitarity in the Minkowskian theory. For instance, $\langle  \phi(-\tau,\bn)\phi(\tau,\bn)\rangle_\text{cyl}=\langle  \phi(\tau,\bn)^\dagger\phi(\tau,\bn)\rangle_\text{cyl}$ corresponds to the (positive)  norm of the state $\phi(\tau,\bn)|0\rangle_\text{cyl}$.
 
When we go back to the flat space, the transformation $\theta$ maps to the inversion $R:x\to x/x^2$. We conclude that \emph{a flat space correlator 
with operators inserted in an inversion-invariant way computes the norm of a state in  radial quantization.} Similarly, when dealing with operators, hermitean conjugation  corresponds to the inversion $R$. 
 For real primary scalars, eqs.~(\ref{eq:cylrd},\ref{eq:hamiltonian_conjugation}) then imply
\beq
\phi(x)^\dagger=r^{-2\Delta_\phi}\phi(Rx)= R(\phi(x))\,.
\eeq
(Analogous rules can be written for operators of nonzero spin.) For example, an inversion-invariant two point function can now be written as
\beq
\langle 0|\phi(x)^\dagger \phi(x)|0\rangle\,,
\eeq
where $|0\rangle$ and $\langle 0|$ are the radial quantization \emph{in} and \emph{out} vacua.

Hermitean conjugation in the Euclidean theory extends to the conformal algebra generators. In particular, we have
\beq
P_\mu^\dagger=K_\mu\,.
\label{eq:PK}
\eeq
This follows from the fact that a special conformal transformation is obtained from a translation by conjugating with the inversion: $K_\mu=RP_\mu R$. One way to check this result is by considering the corresponding differential operators
expressed in terms of the cylinder coordinates:
\begin{align}
P_\mu&=-i\del_\mu\to-ie^{-\tau}[\bn_\mu \del_\tau+(\delta_{\mu\nu}-\bn_\mu\bn_\nu)\del/\del\bn_\nu]\,,\nn\\
K_\mu&=-i[x^2\del_\mu-2x_\mu (x\del)]\to-ie^{\tau}[-\bn_\mu \del_\tau+(\delta_{\mu\nu}-\bn_\mu\bn_\nu)\del/\del\bn_\nu]\,.
\end{align}
We now see that they go one into the other under the time reflection. Moreover when going back to the Minkowskian cylinder, $\tau \to i t$, we do find $K_\mu = P_\mu^\dagger$, implying the same relation for the Hilbert space operators.\footnote{To avoid possible confusion, we should stress that the relation $K_\mu = P_\mu^\dagger$ is only valid in the radial quantization, i.e.~on the cylinder (Euclidean or Minkowski). For the original theory on flat Minkowski space the conformal generators are hermitean: $K^\dagger=K$, $P^\dagger=P$.} 

Relation \reef{eq:PK} has several consequences. First, given a radial quantization state $|\Phi\rangle$ and a conjugate state $\langle\Phi|$ the states 
\beq
P_\mu|\Phi\rangle\qquad\text{and}\qquad \langle\Phi|K_\mu
\eeq
will be also conjugate, as more general states obtained by acting with several $P$'s and $K$'s. Their scalar product will be positive as a norm should.

The second application concerns expressing the operators $\phi(x)$ and $\phi(x)^\dagger$ in terms of the radial quantization ``asymptotic states", i.e.~operators inserted at $0$ and $\infty$. We have the usual QFT formula
\beq
\phi(x)=e^{iP\cdot x}\phi(0)e^{-iP x}
\quad\Longrightarrow\quad \phi(x)|0\rangle=e^{iP\cdot x}\phi(0)|0\rangle\equiv e^{iP\cdot x}|\phi\rangle\,,
\eeq
where we introduced the notation for the state produced by acting with $\phi(0)$ on the \emph{in} vacuum. On the other hand for $\phi(x)^\dagger$ we have the following representation:
\beq
\phi(x)^\dagger=e^{iK\cdot x}\phi(\infty)e^{-iK\cdot x}\quad\Longrightarrow\quad 
\langle 0|\phi(x)^\dagger=  \langle \phi|e^{-iK\cdot x}\,,
\eeq
where the conjugate state $\langle\phi|$ is obtained by acting with $\phi(\infty)$ on the \emph{out} vacuum. The definition of $\phi(\infty)$ as a conjugate of $\phi(0)$ contains as usual a rescaling factor necessary to get a finite limit:  $\phi(\infty)=\lim_{r\to\infty} r^{2\Delta_\phi} \phi(x)$.

\subsection{Reproducing two point function}

We would now like to illustrate the operator formalism with a few simple computations, checking that it gives results agreeing with other techniques.

Let's begin with the scalar two point function. According to the above discussion, it can be evaluated in radial quantization as:
\beq
\langle \calO(x_2) \calO(x_1)\rangle =r_2^{-2\Delta} \langle 0|\calO(R x_2)^\dagger \calO(x_1)|0\rangle =
r_2^{-2\Delta} \langle \calO| e^{-iK\cdot (Rx_2)}  e^{iP\cdot  x_1}|\calO\rangle\,.
\label{eq:lastline}
\eeq
Expanding the exponentials, we get a sum of the same form as \reef{eq:Geg} with
\begin{gather}
c_n=\frac{1}{(n!)^2}\langle n,\bn_2|n,\bn_1\rangle\,,
 \qquad |n,\bn_1\rangle=(P\cdot \bn_1)^n|\calO\rangle\,,\qquad  \langle n,\bn_2|=\langle\calO|(K\cdot \bn_2)^n\,.
 \label{eq:cn1}
 \end{gather}
These are precisely the states mentioned in \reef{eq:states}. The scalar products can be now evaluated using the conformal algebra commutation relations:
\begin{align}
[K_\mu,P_\nu]&=2i(\delta_{\mu\nu} D+M_{\mu\nu})\,,\nn\\
[M_{\mu\nu},X_\rho]&=-i(\delta_{\mu\rho}X_{\nu}-\delta_{\nu\rho}X_{\mu})\qquad(X=P,K)\,,\\
[D,P_\mu]&=-iP_\mu\,,\quad[D,K_\mu]=iK_\mu\,,\nn
\end{align}
taking into account the action of the generators on the primary scalar state $|\calO\rangle$:
\beq
K_\mu|\calO\rangle=M_{\mu\nu}|\calO\rangle=0,\qquad D|\calO\rangle =-i\Delta|\calO\rangle\,.
\eeq
The states $|n,\bn_i\rangle$ with different $n$'s are mutually orthogonal.

It is not difficult to show that $c_n$'s satisfy the Gegenbauer polynomial recursion relation
\beq
n c_n=2(n+\Delta-1)(\bn_1\cdot \bn_2) c_{n-1}-(n+2\Delta-2) c_{n-2}
\eeq
and therefore coincide with \reef{eq:gegexpl}. For the first few terms the agreement is easy to check directly:
\beq
c_0=\langle\Delta|\Delta\rangle=1\,,\quad c_1=2\Delta (\bn_1\cdot\bn_2)\,,\quad c_2=2\Delta(\Delta+1)(\bn_1\cdot\bn_2)^2-\Delta\,,\ldots
\label{eq:firsttwo}
\eeq
Notice that for $\bn_1=\bn_2$ the states $|n,\bn_1\rangle$ and $ \langle n,\bn_2|$ become conjugate. The scalar product $c_n$ then computes their norm and must be positive. Eq.~\reef{eq:firsttwo} is consistent with this.

\subsection{OPE and three point function}

A similar check can be done for the three point function. For simplicity we consider the case $y=0$. Eq.~\reef{eq:3ptsc} then predicts the following radial quantization matrix element:
\begin{align}
\langle 0|\calO(x_2)\phi(x_1)|\phi\rangle
=f_{\phi\phi\calO} r_2^{-2\Delta_\calO} r_1^{\Delta_\calO-2\Delta_\phi}|1-2e^{-\T}(\bn_1\cdot\bn_2)+e^{-2\T}|^{-\Delta_\calO/2}\,.
\label{eq:3ptlargeT}
\end{align}
If needed, this can be also expanded into powers of $e^{-\T}\equiv r_1/r_2$ with Gegenbauer polynomials as coefficients. 

We would like to see how the same matrix element is computed in the Hamiltonian language. 
The key is to interpret the OPE \reef{eq:cOPE},\reef{eq:3t} as an expansion of the state $\phi(x)|\phi\rangle$ into 
energy eigenstates on the cylinder, as we already remarked in section \ref{sec:pol}. Replacing
$\del_y\to -iP$,
we get
\beq
\phi(x_1)|\phi\rangle={\sum_\calO} f_{\phi\phi\calO}r_1 ^{\Delta_\calO-2\Delta_\phi}\Bigl[1+\frac i2 r_1 P\cdot\bn
-r_1^2\{\alpha (P\cdot \bn)^2+\beta P^2\}+\ldots\Bigr]|\calO\rangle\,.
\label{eq:3tcyl}
\eeq
For the bra part $\langle 0|\calO(x_2)$ we use the same expression as in \reef{eq:lastline}.

The three point function computation is now reduced to evaluating matrix elements of products of $K$'s and $P$'s, just like in the two point function case. Using the values for $\alpha$ and $\beta$ given in \reef{eq:3t},
it is easy to check explicitly that the first few coefficients of the $e^{-\T}$ expansion agree with \reef{eq:3ptlargeT}. 

\section{Four point function}
\label{sec:4pt}

In the previous section we saw how the state-operator correspondence can be used to compute two and three point functions.
Since the form of these correlators is anyway fixed by conformal invariance, the point of this exercise was mostly to check the formalism. We will now apply these methods to the OPE convergence argument from section \ref{sec:pol}. 

\subsection{Convergence}
\label{sec:4pt-1}
Consider then the simplest higher point function: the correlator of four identical scalars. There are many ways to map this correlator to the cylinder, depending on the choice of the radial quantization origin and the infinity. For simplicity we will first assume that two out of the four $\phi$ operators are inserted at these two special points: $x_1=0$, $x_4=\infty$. A more symmetric configuration will be useful later on.

In this limit the four point function \reef{eq:4pteq} maps to the radial quantization matrix element
\beq
\langle \phi|\phi(x_3) \phi(x_2)|\phi\rangle\,.
\label{eq:corr4}
\eeq
We wish to compute this matrix element as follows. First we expand the state $\phi(x_2)|\phi\rangle$ by using the OPE in the $\phi(x_2)\times\phi(x_1)$ channel, as in Eq.~\reef{eq:3tcyl}. Second, by conjugating the same equation, we get an analogous expansion of the state
$\langle \phi|\phi(x_3)$ into conjugate states generated by the action of $K$'s. Taking the scalar product, we get a sum of the form:
\beq
{\sum_\calO}(f_{\phi\phi\calO})^2 \sum_{n=0}^\infty e^{-E_n \T}\langle\calO, n,\bn_3|\calO,n,\bn_2\rangle\,,\qquad E_n=\Delta_\calO+n\,,.
\label{eq:OPEdeccyl}
\eeq
Here $|\calO,n,\bn_2\rangle$ are states generated by acting on $|\calO\rangle$ with $n$ powers of $P$ with various contractions involving $\bn_2$, while $\langle\calO, n,\bn_3|$ are the conjugate states involving $K$'s and $\bn_3$. For operators $\calO$ which are Lorentz tensors these states also carry spin indices which are left implicit. The exact structure of these states is fixed by the functional form of the OPE coefficients, as demonstrated in Eq.~\reef{eq:3tcyl} for the scalar $\calO$ case, but we will not need to know it. 

The expansion parameter in \reef{eq:OPEdeccyl} is related to the cylinder time interval:
\beq
e^{-\beta}=r_2/r_3\quad\Leftrightarrow\quad \beta=\tau_3-\tau_2\,.
\eeq
According to the criterion \reef{eq:rad}, the series is expected to converge whenever $\T$ is positive. This looks plausible as the terms with large dimensions are exponentially suppressed for $\T>0$. However, what about the coefficients multiplying the exponentials? Can't they somehow overcome the suppression? Also, shouldn't one worry about the density of states? 

As we will now explain, convergence can be proven in a way which makes no assumptions about the size of the OPE coefficients or the density of states. To do this, consider the matrix element \reef{eq:corr4} as a scalar product of two states
\beq
|\Phi\rangle=\phi(x_2)|\phi\rangle\qquad\text{and}\qquad \langle\Psi|=\langle\phi|\phi(x_3)\,.
\eeq
The norms of these states represent the correlators of the same form as \reef{eq:corr4} but at special, inversion symmetric, configurations. (If $\beta>0$, then by dilatation invariance we can assume that $r_2<1<r_3=r_2^{-1}$ so that the matrix elements defining the norms are radially ordered.) So these norms must be finite (and positive). In fact they can be computed by setting $\bn_2=\bn_3$ in \reef{eq:OPEdeccyl}, which makes every single term positive, being a scalar product of two conjugate states. 

Now, a basic theorem about Hilbert spaces says that the scalar product of two finite norm states converges.
The proof is by judicious application of the Cauchy inequality. We could apply it to the scalar product and bound it by the product of the norms. However, this would not prove that the scalar product \emph{converges}. For this we should consider the tail of the scalar product series \reef{eq:OPEdeccyl} and show that it tends to zero as the cutoff $E_n\ge E_*$ is sent to infinity. By Cauchy, such a tail is bounded by the product of the corresponding tails of the norm series. Since the latter tails consist of positive terms, they \emph{must} tend to zero if the norms are to be finite. Q.E.D.

With convergence of series \reef{eq:OPEdeccyl} established, the next step is to study the convergence \emph{rate}.

\subsection{Spectral density estimates}
\label{sec:est}

 It will be convenient to write the series as the Laplace transform of a ``weighted spectral density"
\beq
\mathcal{L}(\T)=\int_0^\infty dE\,f(E)e^{-E\T}\,,\qquad f(E)\equiv \sum_k\rho_k\, \delta (E-E_k)\,.
\label{eq:Laplace}
\eeq
Here the energies $E_k$ run over all the states present in the theory, primaries or not. The weights $\rho_k$ can be read off from \reef{eq:OPEdeccyl}. We will be considering the reflection positive case $\bn_2=\bn_3$ for which all of these coefficients are positive. As explained above, the general case reduces to this one via the Cauchy inequality.

For each fixed $\beta> 0$, the convergence rate of this integral is clearly controlled by how fast the spectral density grows at large $E$. We therefore begin by studying this second question.

The key observation which will allow us to control the spectral density is that for $\beta\to0$ the four point function \reef{eq:corr4} can be computed by using the leading OPE for the product of two colliding operators $\phi(x_3) \phi(x_2)$. Thus the small $\beta$ behavior of the Laplace transform is fixed:\footnote{Notation $a \sim b$ means $a/b\to 1$ in the assumed limit.}
\beq
\mathcal{L}(\T)\sim \T^{-2\Delta_\phi}\qquad(\beta\to0)\,.
\label{eq:OPEasymp}
\eeq
This asymptotics puts a strong constraint on the behavior of spectral density at large $E$. Were we to assume a power law growth of $f(E)$, Eq.~\reef{eq:OPEasymp} would immediately fix the exponent and the prefactor:
\beq
f(E)\sim \frac{E^{2\Delta_\phi-1}}{\Gamma(2\Delta_\phi)}\qquad \text{(naive)}\,.
\eeq
This of course cannot be interpreted literally as $f(E)$ is a sum of $\delta$-functions. But the integrated 
version of this equation is mathematically true: the integrated spectral density
\beq
F(E)\equiv \int_0^E dE'\,f(E')\
\eeq
must behave asymptotically as
\beq
F(E)\sim\frac{E^{2\Delta_\phi}}{\Gamma(2\Delta_\phi+1)}\,.
\label{eq:FE}
\eeq
This is known as the Hardy-Littlewood tauberian theorem and is surprisingly subtle to prove, if one wants to get the exact prefactor; see \cite{Korevaar}, Theorem I.7.4 for integer-spaced spectra and Theorem I.15.3 for the general case. The difficulty comes precisely from the fact that one does not make any regularity assumption about the weights $\rho_k$ apart from them being positive. 

Denote by $E_{\text{HL}}$ the energy above which the Hardy-Littlewood asymptotics \reef{eq:FE} becomes effective, in the sense that the ratio $F(E)/E^{2\Delta_\phi}$ can be bounded from above and below by positive constants. The precise value of $E_{\text{HL}}$ of course depends on these constants but we will not keep track of this dependence. Analogously suppose that the Laplace transform asymptotics \reef{eq:OPEasymp} is effective for $\beta\le\beta_0$. By dimensional reasoning, we expect $E_{\text{HL}}\approx 1/\beta_0$. An elementary proof of this relation can be given as follows.

\noindent\line(1,0){466}

{\small 
Start with the obvious estimate:
\beq
\mathcal{L}(\beta)\ge \int_0^E dE'\,f(E')e^{-E'\T}\ge e^{-E\T} F(E)\,.
\eeq
This is true for any $\beta$. Fixing $\beta=b/E$ with an arbitrary $b=O(1)$ we get 
\beq
F(E)\le e^b \mathcal{L}(b/E)\le\emph{const.}\, e^b b^{-2\Delta_\phi} E^{2\Delta_\phi}.
\label{eq:elem}
\eeq
The second inequality is true as long as $b/E\le \beta_0$, i.e.~$E\ge E_\text{HL}=b/\beta_0$. A lower bound on $F(E)$ can also be shown by similar tricks; see \cite{Titchmarsh}, Section 7.5.2.

In the asymptotic limit $E\to\infty$ we could take $const.\to1$ in the last inequality. The prefactor can then minimized by picking $b=2\Delta_\phi$,  but it is still worse than the one in \reef{eq:FE}. To get the optimal prefactor, a much more elaborate argument is required.
}

\noindent\line(1,0){466}

Finally, let us discuss the error term estimates in the Hardy-Littlewood asymptotics. The error in \reef{eq:OPEasymp} is controlled by $\Delta_0$, the dimension of the first non-trivial operator in the $\phi\times\phi$ OPE:
\beq
\mathcal{L}(\T)= \T^{-2\Delta_\phi}[1+O(\T^{-\Delta_0})]\,.
\label{eq:OPEasymp0}
\eeq
One could wonder if this power-suppressed error would translate into a similarly power-suppressed error term in \reef{eq:FE}. However, this expectation is wrong, proving once more that this theorem is more subtle than it seems. In fact, the best possible error estimate under these conditions is only logarithmic (independently of $\Delta_0$):
\beq
\reef{eq:OPEasymp0}\quad\Longrightarrow\quad  
F(E)=\frac{E^{2\Delta_\phi}}{\Gamma(2\Delta_\phi+1)}[1+O(1/\log E)]\,.
\label{eq:remainder}
\eeq
See \cite{Korevaar}, Chapter VII, for this result and for the elementary examples showing that the estimate cannot be improved without further assumptions. 

\subsection{Comparison with partition function spectral densities}

To appreciate better the estimate \reef{eq:FE}, let us reproduce the parallel story for the usual, ``unweighted", spectral density
\beq
f_0(E)\equiv \sum_k \delta (E-E_k)\,,
\eeq
which counts all the states with equal weight one. In this case the Laplace transform $\mathcal{L}_0(\beta)$ has the interpretation of the CFT partition function on $S^{D-1}$ times a circle of radius $\beta$. The $\beta\to0$ asymptotics then takes quite a different form, as we expect
\beq
\mathcal{L}_0(\beta)\sim \exp[-\text{vol}(S^{D-1})\,\mathcal{F}(\beta)]\qquad(\beta\to0)\,,
\eeq
where $\mathcal{F}(\beta)$ is the CFT free energy density in flat space $\bR^{D-1}\times$ time. This is because in the high-temperature limit the curvature of $S^{D-1}$ can be neglected.

Furthermore, by dimensional analysis the CFT free energy density must be given by (see e.g.~\cite{ElShowk:2011ag}):
\beq
\mathcal{F}(\beta)=-\kappa/\beta^{D-1}\,.
\eeq
The constant $\kappa>0$ provides a measure of the number of degrees of freedom and one expects it to be $O$(central charge). In 2D, this can be shown rigorously using modular invariance:
\beq
\kappa=(\pi/12)c\,\qquad\text{(2D)}.
\label{eq:2Dkappa}
\eeq

So, $\mathcal{L}_0(\beta)$ grows exponentially fast as $\beta\to0$, and this requires an exponential growth of the corresponding spectral density. Using the stationary phase approximation, we find:
\beq
f_0(E)\approx \exp[\gamma E^{1-\frac{1}{D}}]\,,\qquad \gamma=\frac{D}{D-1}\left[\frac{\text{vol}(S^{D-1})}{D-1}\,\kappa\right]^{1/D}\,.
\label{eq:f0}
\eeq 
This should be understood as an order of magnitude estimate after averaging; see \cite{Korevaar}, Section IV.16, for a rigorous formulation. For $D=2$ and $\kappa$ as in \reef{eq:2Dkappa}, this result reduces to the well-known formula for the asymptotic density of states in 2D CFTs.

Eq.~\reef{eq:f0} puts our result \reef{eq:FE} into nice perspective: the unweighted CFT spectral density entering the partition function grows exponentially fast, but the weighted ones relevant for the correlator computations have much slower power-like growth. This means that the OPE coefficients squared must be exponentially suppressed.\footnote{{\bf Note added (August 2014):} This statement is supposed to be true only on average. For example, in free field theory, in the OPE $\phi\times\phi$ only a tiny fraction of all operators appears with nonzero coefficients (one operator per spin), and the coefficients of these operators are therefore not exponentially small. 
In a general CFT, most of the exponentially many operators are expected to appear with nonzero (hence exponentially small) coefficients, but there are always special series of operators, whose coefficients are not exponentially suppressed \cite{Fitzpatrick:2012yx,Komargodski:2012ek} .} 
\subsection{Convergence rate}

We will now study the convergence rate of the OPE representation \reef{eq:OPEdeccyl}, written as the Laplace transform \reef{eq:Laplace}. In other words, we must study how fast the tail of this integral
\beq
\mathcal{L}(\T,E_*)=\int_{E_*}^\infty dE\,f(E)e^{-E\T}\,.
\label{eq:OPEtail}
\eeq
tends to zero for $E_*\to\infty$ and a fixed $\beta>0$. 

The necessary estimation is done as follows:
\begin{align}
\mathcal{L}(\T,E_*)&=\beta\int_{E_*}^\infty dE\,[F(E)-F(E_*)]e^{-E\T}\le \beta \int_{E_*}^\infty dE\,F(E)\,e^{-E\T}\nn\\
&\sim\beta \int_{E_*}^\infty dE \frac{E^{2\Delta_\phi}}{\Gamma(2\Delta_\phi+1)} \,e^{-E\T}=\beta^{-2\Delta_\phi}\frac{\Gamma(2\Delta_\phi+1,E_*\beta)}{\Gamma(2\Delta_\phi+1)}\,.
\end{align}
Here in the first line we integrated by parts and then dropped the $F(E_*)$ term; keeping it is not useful because of the very weak error estimate in \reef{eq:remainder}. We then used the asymptotics \reef{eq:FE} and wrote the answer in terms of the incomplete gamma function.

The most interesting case of this estimate occurs for $E_*\gg \Delta_\phi/\beta$. Using the incomplete gamma function asymptotics, we
get\footnote{In this paper $a\lesssim b$ means that $\lim\sup a/b\le 1$ in the considered limit, $E_*\to\infty$ in this particular case.}
\beq
\mathcal{L}(\T,E_*)\lesssim \frac{1}{\Gamma(2\Delta_\phi+1)}E_*^{2\Delta_\phi} e^{-E_*\beta}\qquad (E_*\gg \Delta_\phi/\beta,\ E_*\gg E_\text{HL})\,.
\label{eq:OPEconv}
\eeq
We also added explicitly the condition that $E_*$ should be above $E_\text{HL}$, so that the used spectral density asymptotics be applicable.

Estimate \reef{eq:OPEconv} shows that \emph{the OPE expansion for the four point function converges exponentially fast.} Given the importance of this result, we will give its alternative derivation directly from \reef{eq:OPEasymp}, bypassing the spectral density estimates. 

\noindent\line(1,0){466}

{\small 
 We can write:
\beq
\mathcal{L}(\T,E_*)= \int_{E_*}^\infty dE\,f(E)e^{-E\beta'} \times e^{-E(\beta-\beta')},
\eeq
where $\beta'\in(0,\beta)$ is a parameter to be fixed below.
Now estimate the last factor by $e^{-E_*(\beta-\beta')}$. The remaining integral is $\mathcal{L}(\T', E_*)$ but estimate it generously by $\mathcal{L}(\T')$. We obtain
\beq
\mathcal{L}(\T,E_*)\le \mathcal{L}(\T') e^{-E_*(\beta-\beta')} \le const.\, (\beta')^{-2\Delta_\phi}e^{-E_*(\beta-\beta')}\,,
\eeq
as long as $\beta'$ is smaller than $\beta_0$ introduced in section \ref{sec:est}, so that the asymptotics \reef{eq:OPEasymp} is effective. The $\beta'$ dependence is minimized for 
\beq
\beta'=2\Delta_\phi/E_*\,,
\label{eq:beta'}
\eeq
and we get
\beq
\mathcal{L}(\T,E_*)\le  C E_*^{2\Delta_\phi}e^{-E_*\beta}\,
\eeq
where $C=const.(e/2\Delta_\phi)^{2\Delta_\phi}$ is the same sub-optimal constant that we had in \reef{eq:elem}. For the estimate to be valid, we must have that $\beta'$ defined via \reef{eq:beta'} be smaller than both $\beta$ and $\beta_0$, which requires
\beq
E_*\ge 2\Delta_\phi/\beta,\ E_*\ge 2\Delta_\phi/\beta_0\,.
\label{eq:cutoff}
\eeq
These are the analogues of the two conditions present in Eq.~\reef{eq:OPEconv}.
}

\section{Conformal block decomposition}
\label{sec:cb}

In the previous section we discussed convergence of the OPE series representation of a four point function. This series \reef{eq:OPEdeccyl} was a double series, the first summation being over primaries appearing in the OPE and the second over the descendants of each primary. As a matter of fact, this second sum over descendants computes the conformal block of a given primary:
\beq
G_{\calO}(u,v)=\sum_{n=0}^\infty e^{-E_n \T}\langle \calO, n,\bn_3|\calO,n,\bn_2\rangle\,.
\label{eq:GPos}
\eeq
The values of $u,v$ are found by taking the limit $x_1\to0$, $x_4\to\infty$ in their defining equations. We find:
\beq
u=(r_2/r_3)^2,\qquad v=1+(r_2/r_3)^2-2(r_2/r_3)\cos\alpha\,,
\eeq
where $\alpha$ is the angle between $\bn_2$ and $\bn_3$. It is then easy to solve for $z$ defined in \reef{eq:zzbar}:
\beq
z=(r_2/r_3)e^{i\alpha}\,.
\eeq

It's interesting to compare expression \reef{eq:GPos} with the standard definition of conformal blocks given in \reef{eq:cbdef}. These expressions are not manifestly identical, for the following reason. In \reef{eq:cbdef} the usual OPE is used on both sides of the four point function. On the other hand, when deriving \reef{eq:GPos} we use the OPE on the RHS of the correlator, and the \emph{conjugate OPE} on the LHS. One consequence of this difference is that expression \reef{eq:GPos} is term-by-term positive for real $0<z<1$, which corresponds to reflection positive point configurations, while the definition \reef{eq:cbdef} is not manifestly positive. In spite of these apparent differences, the conformal blocks defined by \reef{eq:GPos} of course agree with the old definition as they should. As a check, we can evaluate the first few terms of small $z$ expansion for the case of scalar $\calO$, using the OPE coefficients from \reef{eq:3tcyl} and norms from \reef{eq:firsttwo}. Focussing on real $z$ for simplicity, we find agreement with the formula
\beq
G_{\Delta,0}(z,z) =
\left(\frac{z^{2}}{1-z}\right)^{\Delta/2}
\, _3F_2\left({\textstyle\frac{\Delta}2,\frac{\Delta}2,\frac{\Delta-D
   }{2}+1 ;\frac{\Delta+1 }{2},\Delta -\frac{D}{2}+1}
   ;\frac{z^2}{4 (z-1)}\right)\,
\eeq
recently derived in \cite{ElShowk:2012ht}.

Notice that representation \reef{eq:GPos} makes obvious two other general properties of the conformal blocks: they \emph{increase} for real $0<z<1$ and \emph{decrease} as $z$ rotates away from the real axis. More precisely, the second property reads
\beq
|G_{\calO}(z=r e^{i\alpha},\bar z=r e^{-i\alpha})|\le G_{\calO}(r,r) \qquad(0<r<1)\,;
\label{eq:zrot}
\eeq
it follows by Cauchy inequality.

\subsection{Convergence}

We now turn to the question posed in the introduction: what is the region and rate of convergence of the conformal block decomposition \reef{eq:CBdeceq}? By convergence we mean convergence of the partial sums
\beq
\lim_{\Delta_*\to\infty}\sum_{\Delta_\calO\le\Delta_*} f_\calO^2\, G_\calO(u,v)\,,
\eeq
and by convergence rate the speed at which the tails
\beq
g_{\Delta\ge\Delta_*}(u,v)=\sum_{\Delta_\calO\ge\Delta_*} f_\calO^2\, G_\calO(u,v)\,,
\eeq
tend to zero.

These questions can be answered based on the results of section \ref{sec:4pt}. The point is that the tails of the conformal block series include but a subset of the terms in the tails \reef{eq:OPEtail} of the OPE series, identifying $\Delta_*=E_*$. This means that the conformal block decomposition converges not worse than the OPE series.\footnote{To be precise, this argument should first be done for real $z$, where the series are term-by-term positive, and then extended to complex $z$ by Cauchy inequality.} Applying the main result \reef{eq:OPEconv} about the OPE convergence, we conclude that the conformal block decomposition converges exponentially fast for each $z$ in the unit disc $r=|z|<1$ with the tail estimate:
\beq
|g_{\Delta\ge\Delta_*}(z,\bar z)|\lesssim \frac{1}{\Gamma(2\Delta_\phi+1)}\Delta_*^{2\Delta_\phi} r^{\Delta_*}\quad\text{for}\quad \Delta_*\gg \Delta_\phi/(1-r)\,.
\label{eq:CBconv}
\eeq

\subsection{Extending the region of convergence}
\label{sec:ext}

Eq.~\reef{eq:CBconv} is sufficient to prove convergence of the conformal block decomposition at $z=1/2$, the point used in the conformal bootstrap applications described in the Introduction. However, this is not the best possible estimate of the convergence rate. This can already be suspected from the fact that conformal blocks are known to be real-analytic in the full $z$ plane minus the cut $(1,+\infty)$ along the positive real $z$ axis. We will now show that the conformal block decomposition converges in this larger region, with the following rate:$^{\ref{note:2015}}$
\beq
|g_{\Delta\ge\Delta_*}(z,\bar z)|\lesssim \frac{1}{\Gamma(4\Delta_\phi+1)}\Delta_*^{4\Delta_\phi} |\rho(z)|^{\Delta_*}.
\label{eq:CBconv1}
\eeq
Here the function 
\beq
\rho(z)= \frac{z}{(1+\sqrt{1-z})^2}
\label{eq:rz}
\eeq
maps the cut plane on the unit disc. Thus the convergence is exponentially fast everywhere in the cut plane. For $|z|<1$ we have  $|\rho(z)/z|<1$ and so the rate of exponential convergence predicted by \reef{eq:CBconv1} is stronger than the one following from \reef{eq:CBconv}.
For $z=1/2$ we have $\rho(z)\approx 0.17$. 

As a matter of fact, the same function $\rho(z)$ enters the individual conformal blocks asymptotics which can be derived \cite{Rattazzi:2008pe} in $D=2$ and $D=4$ starting from the explicit formulas of \cite{Ferrara:1974ny,DO1,DO2}. While this by itself would not be enough to prove \reef{eq:CBconv1}, it hints at the best possible character of this estimate. As we discuss below, this is indeed the case.

To show the new estimate, we will have to change the quantization origin. To compute conformal blocks at $|z|>1$ in radial quantization as an overlap of two states,
we need to find a sphere separating $x_1$ and $x_2$ from $x_3$ (and $x_4=\infty$). Such a sphere exists for all $z$ in the cut plane, but for $|z|>1$ it can no longer be chosen centered at $x_1$.

\begin{figure}[htbp]
\begin{center}
\includegraphics[scale=1]{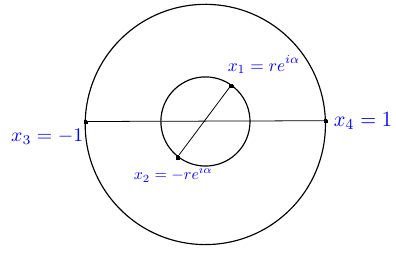}
\caption{Here we are considering four operators inserted at the shown points in a plane passing through the origin. We give complex coordinates in this plane. The circles are intersections of the spheres of radius 1 and $r$ with the plane. The angle $\alpha$ is arbitrary.}
\label{fig:newconf}
\end{center}
\end{figure}

\begin{figure}[htbp]
\begin{center}
\includegraphics[scale=1]{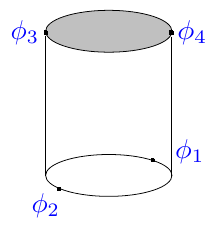}
\caption{The configuration on $S^{D-1}\times \bR$ obtained from Fig.~\ref{fig:newconf} by the Weyl transformation. The pairs of points $\phi_{3,4}$ and $\phi_{1,2}$ are in antipodal positions on the spheres at the cylinder time $0$ and $\log r$. Their positions on their respective spheres are rotated with respect to each other by angle $\alpha$.}
\label{fig-cyl}
\end{center}
\end{figure}

The needed geometry is best captured by inserting the four operator as in Fig.~\ref{fig:newconf}.
On the cylinder, this configuration maps into the one shown in Fig.~\ref{fig-cyl}. Here $\phi_{3,4}$ are inserted at the cylinder time $0$, while $\phi_{1,2}$ at smaller time $\log r<0$. Computing the cross ratios and the value of $z$ in this geometry, we find
\beq
z={4 \rho}/{(1+\rho)^2}\qquad(\rho\equiv r\,e^{i\alpha})\quad\Leftrightarrow\quad \rho=\rho(z)\,.
\label{eq:zr1}
\eeq
In other words, $z(\rho)$ is precisely the inverse of the function \reef{eq:rz}.

We now have to repeat the main steps leading to \reef{eq:CBconv} in the new geometry. First of all we should express the radial quantization matrix element
\beq
\langle 0|\phi(x_3)\phi(x_4)\phi(x_1)\phi(x_2)|0\rangle
\eeq
using the OPE with respect to the new \emph{in} and \emph{out} vacuum positions. To begin with, the state $\phi(x_1)\phi(x_2)|0\rangle$ is expanded into states generated by primaries and their descendants inserted at the origin. This expansion represents an equivalent form of OPE, the difference from \reef{eq:cOPE} being that we have $\calO(\frac{x+y}2)$ and not $\calO(y)$ in the RHS. It's trivial to work out coefficient functions in such modified OPE in terms of those appearing in \reef{eq:cOPE}. We will not need their exact form below. Schematically we will get:
\beq
\phi(x_1)\phi(x_2)|0\rangle = (2r)^{-2\Delta_\phi}{\sum_\calO}f_{\phi\phi\calO} \sum_{n=0}^\infty r^{\Delta_\calO+n}|\calO,n,\alpha\rangle,
\eeq
where the $n$ sum is over descendants, the states $|\calO,n,\alpha\rangle$ being generated by acting on $|\calO\rangle$ with $n$ powers of $P_\mu$ in various contractions. Apart from the overall $f_{\phi\phi\calO}$ factors, their form and the relative coefficients are completely fixed by conformal symmetry. In the considered geometry these states also have dependence on the angle $\alpha$ which we don't need to know explicitly.

Conjugating the last equation, we can write an expansion of the state $\langle0|\phi(x_3)\phi(x_4)|$ in terms of the states $\langle\calO,n,\alpha=0|$ obtained by acting with $K$'s on the primary out state $\langle \calO|$. Taking the overlap of the two expansion, the four point function on the cylinder is expressed as the double series
\beq
{\sum_\calO}(f_{\phi\phi\calO})^2 \sum_{n=0}^\infty r^{\Delta_\calO+n}\langle\calO,n,0|\calO,n,\alpha\rangle.
\label{eq:newser}
\eeq
By the same argument as in section \reef{sec:4pt-1}, we conclude that this series converges for all $r<1$.

To estimate the convergence rate we go to the reflection positive case $\alpha=0$ and consider the limit $r\to1$. In this limit the sum of the series \reef{eq:newser} must behave consistently with the OPE. Two pairs of the operators are colliding simultaneously, so the asymptotics is given by $\sim (1-r)^{-4\Delta_\phi}$. This in turn implies that the integrated spectral density must behave asymptotically as\footnote{\label{note:2015}{\bf Note added (June 2015):} Eqs.~\reef{eq:CBconv1}, \reef{eq:F(E)}, \reef{eq:OPEconv1} should have an extra factor $2^{4\Delta_\phi}$ inserted in the RHS. When the OPE asymptotics of the 4pt function for $r\to1$ is translated into the asymptotics of the function $g(u,v)$, one should remember about the denominator in \reef{eq:4pteq}. In the geometry of Fig.~\ref{fig:newconf}, this factor is $(2r)^{2\Delta_\phi}  2^{2\Delta_\phi}\to 2^{4\Delta_\phi}$ as $r\to 1$. We are grateful to Marco Serone for pointing out some paradoxes which helped us find the extra factor. Notice that in the geometry relevant for section 5.1, where points are inserted at $0,z,1,\infty$, the corresponding factor is $|z|^{2\Delta_\phi}\to 1$ $(z\to 1)$. Thus \reef{eq:CBconv} is valid as written without any extra factors. }
\beq
F(E)\sim \frac{E^{4\Delta_\phi}}{\Gamma(4\Delta_\phi+1)}\,.
\label{eq:F(E)}
\eeq
By analogy with Eq.~\reef{eq:OPEconv}, the tails of the sum \reef{eq:newser} obtained by cutting off the energies $E_n=\Delta_\calO+n>E_*$ must therefore decay exponentially as$^{\ref{note:2015}}$
\beq
\lesssim \frac{1}{\Gamma(4\Delta_\phi+1)}E_*^{4\Delta_\phi} r^{E_*}\quad\text{for}\quad E_*\gg \Delta_\phi/(1-r)\,.
\label{eq:OPEconv1}
\eeq
This is the needed estimate, since for $\alpha=0$ we have $r=\rho(z)$ according to \reef{eq:zr1}. We just have to interpret it for the conformal block decomposition. Just as before, conformal blocks are given by the sums over $n$ in \reef{eq:newser}. Conformal block decomposition tails can be simply bounded by the OPE series tails with $E_*=\Delta_*$, since the latter tails include more terms and all the terms in the considered case $\alpha=0$ are positive. 

Finally, consider nonzero $\alpha$. The matrix elements entering the OPE series can be bounded by Cauchy inequality as follows:
\beq
|\langle\calO,n,0|\calO,n,\alpha\rangle|\le \sqrt{\langle\calO,n,0|\calO,n,0\rangle \langle\calO,n,\alpha|\calO,n,\alpha\rangle}= \langle\calO,n,0|\calO,n,0\rangle\,,
\eeq
where we used that the norms in the RHS are equal by rotation invariance on $S^{D-1}$. In particular, individual conformal blocks can only decrease in absolute value as $\alpha$ becomes nonzero keeping $r$ fixed:
\beq
\bigl|G_\calO|_{\alpha\ne0}\bigr|\le G_{\calO}|_{\alpha=0}.
\eeq
This inequality is the analogue of Eq.~\reef{eq:zrot} in the new geometry. Therefore the conformal block decomposition tails at $\alpha\ne0$ satisfy the same estimate as in the already considered case $\alpha=0$. This is precisely what is stated in \reef{eq:CBconv1}, as $r=|\rho(z)|$ according to \reef{eq:zr1}. Q.E.D.

We now explain in which sense Eq.~\reef{eq:CBconv1} is best possible. First of all, this can be checked numerically in simple solvable CFTs. For example, one can consider the four point function of free scalar boson in 4D whose conformal block decomposition is known explicitly \cite{DO1}. Focussing on real $0<z<1$ for simplicity, one finds that the convergence rate is well fit by
\beq
C(z) \Delta_*^{4-\gamma(z)} \rho(z)^{\Delta_*}.
\eeq
Here $\gamma(z)\approx 1.5$ weakly dependent on $z$. This shows that while the power-like prefactor $\Delta_*^{4\Delta_\phi}$ is improvable, the exponential convergence rate is best possible.

It is also instructive to see why it would be impossible to do better than $\rho(z)^{\Delta_*}$ using our method.
Consider the following family of reflection-positive operator insertion configurations:\footnote{A more general family of reflection-positive configurations obtained from \reef{eq:inter}
by rotating $x_1$ and $x_4$ with respect to the origin by the same angle $\varphi$ leads to the same conclusion.}
\beq
x_1=r,\ x_4=1,\ x_2=-kr,\ x_3=-1/k.
\label{eq:inter}
\eeq
For $r<1$ and $k$ between 0 and 1 this family interpolates between the configurations considered in section \ref{sec:4pt} and here.
We also have an interpolating family of four point functions expansions, with the operators in the RHS of the OPE always inserted at the origin and at infinity. The $r\to1$ limit corresponds to colliding operators, and the Laplace transform has a power-like singularity. The order of this singularity will depend on whether $k<1$ (one pair collides) of $k=1$ (two pairs collide). But the leading exponential asymptotics of the Laplace transform tails at finite $r$ will be the same: $r^{\Delta_*}$. Now, the conformal invariant $z$ for the configuration \reef{eq:inter} is given by:
\beq
z=(1+k)^2r/(1+k r)^2\,.
\eeq
To maximize the convergence rate for a given $z$, we must find a $k$ such that the value of $r$ found from this equation is smallest. It is then simple to check that $k=1$ is always the optimal choice. 

\section{Discussion}
\label{sec:disc}

The purpose of this paper was to provide a set of criteria for determining the region and the rate of convergence of the OPE and of the conformal block expansions routinely used in CFT. For simplicity, we focused on the four point function of identical scalars, although the method is general. Our main results are Eq.~\reef{eq:OPEconv} and \reef{eq:OPEconv1}, establishing the exponential convergence rate of the OPE for two different schemes, depending whether $\phi(x)\phi(y)$ is expressed as a sum of operators inserted at $y$ or $(x+y)/2$. Convergence of the conformal block decomposition was shown to follow from that of the OPE. The optimal estimate \reef{eq:CBconv1} is obtained by using the second scheme. 
Near the point $z=1/2$ used in the conformal bootstrap studies, truncating the conformal block decomposition at scaling dimension $\Delta_*\gg 1$ induces the error $\lesssim 0.17^{\Delta_*}$. 

Convergence is good news for the conformal bootstrap, because it means that the method and the results obtained so far are on mathematically solid ground. Fast convergence also means that approximate expressions for the conformal blocks obtained by throwing out the descendants beyond certain level may be used where all-order expressions are not yet easily available, like for external operators with spin. On the other hand, exponentially fast convergence also has a downside: it means that the very high-dimension operators effectively decouple, and it will be very difficult to learn about them by studying consistency conditions for the correlators of low-dimension operators,
how it's done in conformal bootstrap.\footnote{One possibility to enhance the contribution of high-spin operators is to consider consistency conditions for the correlators analytically continued to the Minkowski space.} 

For CFTs with a gravitational dual, decoupling of high-dimension operators can be understood as an ordinary effective field theory decoupling of high-mass states in the bulk. In the AdS/CFT context, a CFT truncated to just the subsector of low-dimension operators has been dubbed ``Effective CFT" \cite{Fitzpatrick:2010zm}. Our results show that such a truncation gives an approximate description in any CFT, with or without AdS dual.

To prove the exponentially fast convergence, we represented the four point function as the Laplace transform of a spectral density concentrated on the sequence of dimensions of the exchanged local operators (primaries and descendants). The weights were given by the OPE coefficients squared times kinematical factors (the descendant norms). The main observation was that the short-distance limit of the Laplace transform could be controlled by using the OPE for the colliding operators. By the Hardy-Littlewood tauberian theorem, this implied a power-like asymptotic behavior for the (averaged) spectral density. From here it was fairly simple to show that the Laplace transform tails \emph{at finite operator separation} are exponentially small.
 
The power-like asymptotics of the weighted spectral density is interesting in its own right. As we reviewed in section \ref{sec:est}, the same density with unit weights (i.e.~simply counting states) is expected to have an exponentially growing asymptotics $\approx \exp[\gamma E^{1-\frac 1D}]$. To make these two statements compatible, the individual OPE coefficients must be exponentially suppressed with the same exponent. 

Recently, \emph{faster-than}-exponential suppression for OPE coefficients of high-dimension $\Delta$ and not-too-high spin $l$ operators was discussed in \cite{Fitzpatrick:2011hu}. For large-$N$ theories with gravity duals, they have an estimate
\beq
f_{\Delta,l}\sim \exp\left[-\frac 12\left(\frac{\Delta^{D-2}}{N^2}\right)^{\frac 1{D-3}}\right]\quad\text{for}\quad\Delta\gg N^{\frac{2}{D-2}},\quad
l \ll \left(\frac{\Delta^{D-2}}{N^2}\right)^{\frac 1{D-3}} \,.
\label{eq:conj}
\eeq
This results follows from the fact that in the limit of large AdS radius the conformal block decomposition of the CFT four point function goes into the partial wave decomposition of the flat space gravitational S-matrix. High-dimension, low angular momentum contributions in \reef{eq:conj} correspond to the transplanckian scattering at small impact parameters. Black hole production is then expected to dominate, while the elastic $2\to2$ amplitude will be exponentially suppressed by the black hole entropy. The corresponding piece of the four point function, $f_{\Delta,l}^2$ times the conformal block, should then be similarly suppressed. 

Clearly, Eq.~\reef{eq:conj} comes from a very different physics input compared to our results. Thus it should not be surprising that it is stronger, predicting faster-than-exponential decay in the given range.\footnote{For this reason it can be stated as a result for $f_{\Delta,l}$ alone, since the conformal blocks behave like simple exponentials.} However, there remain points of contact between the two approaches. For example, it should be possible to show using our methods that, quite generally, the super-exponential suppression cannot extend for all spins (see below). This is also what is expected from the gravitational S-matrix point of view, since for large impact parameters near-elastic $2\to2$ scattering replaces black hole formation.

Our results can be developed in several directions. Finding interesting generalizations involving different external states (unequal operators, operators with spin or with asymptotically growing dimension) is one line of research. A less obvious question is how to make better use of the crossing symmetry constraint \reef{eq:cross-sum}. As in \cite{Rattazzi:2008pe}, let us rewrite it dividing by the contribution of the unit operator:
\beq
\sum_\calO (f_\calO)^2 F_\calO(u,v)\equiv 1,\qquad F_\calO(u,v)=
\frac{v^{\Delta_\phi}G_{\mathcal{O}}(u,v)-u^{\Delta_\phi}G_{\mathcal{O}}(v,u)}
{u^{\Delta_\phi}-v^{\Delta_\phi}}\,.
\label{eq:sumrule}
\eeq
This ``sum rule" has to be satisfied for any $u,v$ where it converges (i.e.~in the cut $z$ plane as explained in section \ref{sec:ext}).
A cornerstone of our argument was Eq.~\reef{eq:OPEasymp}, obtained by using the OPE in the crossed channel. As such it is also a form of crossing symmetry, but a very weak one: it can be interpreted as saying that Eq.~\reef{eq:sumrule} must be satisfied in the limits $z\to0,1$. 

However, Eq.~\reef{eq:OPEasymp} has a shortcoming: it does not say how fast the limit is reached. Suppose that we want to have $\mathcal{L}(\beta)\le 2 \beta ^{-2\Delta_\phi}$ for $\beta<\beta_0$, how small a $\beta_0$ should we take? \emph{A priori} this is not known, the answer may depend on a CFT and on a correlator under consideration. This ignorance feeds into the minimal value of the dimension cutoff above which our exponential asymptotic estimates become rigorous inequalities, as in Eq.~\reef{eq:cutoff}. It would be nice to eliminate this non-constructive feature from our results, by showing that a universal $\beta_0$ can be chosen. Preliminary work \cite{Espin,unpublished} indicates that for $D=4$ this may be done starting from the sum rule \reef{eq:sumrule}, at least for $\phi$'s of dimension less than about 1.7.

The sum rule can be also used to explain why we believe (as mentioned above) that the super-exponential decay of all, rather than just low spin, OPE coefficients would be inconsistent with crossing symmetry. Morally such an assumption would be not too different from throwing out all high dimension primaries from the sum rule. Considering the case of real $0<z<1$, the functions $F_\calO$ are peaked for $z\sim 1-O(1/\Delta^2)$ and go to zero as $z$ approaches 1 past this value. Thus it seems impossible to satisfy the sum rule near $z=1$ under such conditions. It would be interesting to work out the details of this argument.
  
To conclude, we would like to emphasize once more the fundamental principle used in deriving the results of this paper: that the CFT correlation functions can be computed using the quantum-mechanical dynamics on the cylinder $S^{D-1}\times \bR$, which is Weyl equivalent to the flat Euclidean space. It was assumed that we have a complete Hilbert space of normalizable states on the sphere $S^{D-1}$, that the Hamiltonian can be diagonalized, and that the energy eigenstates form a basis. Given these assumptions it is possible to prove the existence of the OPE by expanding the state $\phi(x)\phi(y)|0\rangle$ into the states of fixed energy. It is also a simple matter to convince oneself that any state on the sphere gives rise to a local operator in the flat space (and vice versa), proving the state-operator correspondence.

The discreteness of the spectrum \emph{per se} was not assumed: all of our arguments go through if the sums over primary operators are replaced by integrals. The spectrum of operator dimensions is expected to be discrete in $D\ge 3$. In $D=2$ there exist CFTs with a continuous spectrum of scaling dimensions, such as the free scalar boson with its vertex operators $e^{ipX}$. The Liouville theory and non-compact sigma-models
also have continuous spectrum. In such cases care must be taken in applying our results, since local operators generate states which are not normalizable (they are rather $\delta$-function normalizable). 

As the very last comment, we should quote two monumental axiomatic quantum field theory papers \cite{Luscher:1975js,Mack:1976pa} devoted to the question of OPE convergence in CFT. Most of their efforts seem to go towards establishing rigorously the first step of our argument: that the OPE can be viewed as an expansion in the Hilbert space sense, after which the convergence is automatic. The question of convergence rate was not discussed.

\section*{Acknowledgements}

We are grateful to John Cardy, Sheer El-Showk, Liam Fitzpatrick, Davide Gaiotto, Jared Kaplan, Zohar Komargodski, Hugh Osborn, Kyriakos Papadodimas and Leonardo Rastelli for useful comments and discussions. 
The work of S.R. is supported in part by the European Program ÒUnification in the LHC EraÓ, contract PITN-GA-2009-237920 (UNILHC), and by the \'Emergence-UPMC-2011 research program; he also thanks the Perimeter Institute for hospitality. The work of D.P. and R.R. is supported by the Swiss National Science Foundation under grant 200021-125237.

\bibliography{Biblio-ECFT}{}
\bibliographystyle{utphys}

\end{document}